\documentclass[aps,prb,twocolumn,showpacs,superscriptaddress, preprintnumbers,amsmath,amssymb]{revtex4-1}

\usepackage{graphicx}
\usepackage{epsfig}
\usepackage{dcolumn}
\usepackage{bm}
\usepackage{longtable}
\usepackage{color}

\begin{document}


\title{Multi-band superconductivity in LaFeAsO$_{0.9}$F$_{0.1}$
single crystals probed by high-field vortex torque magnetometry}


\author{G.\ Li}
\affiliation{National High Magnetic Field Laboratory, Florida
State University, Tallahassee-FL 32310, USA}
\author{G.\ Grissonanche}
\affiliation{National High Magnetic Field Laboratory, Florida
State University, Tallahassee-FL 32310, USA}
\author{A.\ Gurevich}
\affiliation{National High Magnetic Field Laboratory, Florida
State University, Tallahassee-FL 32310, USA}
\author{N.\ D.\ Zhigadlo}
\affiliation{Laboratory for Solid State Physics, ETH
Z\"{u}rich, CH-8093 Z\"{u}rich, Switzerland}
\author{S.\ Katrych}
\affiliation{Laboratory for Solid State Physics, ETH
Z\"{u}rich, CH-8093 Z\"{u}rich, Switzerland}
\author{Z.\ Bukowski}
\affiliation{Laboratory for Solid State Physics, ETH
Z\"{u}rich, CH-8093 Z\"{u}rich, Switzerland}
\author{J.\ Karpinski}
\affiliation{Laboratory for Solid State Physics, ETH
Z\"{u}rich, CH-8093 Z\"{u}rich, Switzerland}
\author{L. Balicas} \email{balicas@magnet.fsu.edu}
\affiliation{National High Magnetic Field Laboratory, Florida
State University, Tallahassee-FL 32310, USA}


\date{\today}

\begin{abstract}
To probe manifestations of multiband superconductivity in
oxypnictides, we measured the angular dependence of the magnetic
torque $\tau(\theta)$ in the mixed state of
LaO$_{0.9}$F$_{0.1}$FeAs single crystals as a function of
temperature $T$ and magnetic fields $H$ up to 18 T. The paramagnetic contribution of the Fe ions is properly treated in order to extract the effective mass anisotropy parameter  $\gamma=(m_c/m_{ab})^{1/2}$ from $\tau(\theta)$. We show that  $\gamma$ depends strongly on both $T$ and $H$, reaching a maximum value of $\sim$ 10 followed by a  decrease towards values close to 1 as $T$ is lowered. The observed field dependencies of  the London penetration depth $\lambda_{ab}$ and $\gamma$ suggest the onset of suppression of a superconducing gap at $H \approx H_{c2}/3$.
\end{abstract}


\maketitle


\section{Introduction}
The recently discovered superconducting oxypnictides
\cite{discovery,chen} have similarities with the high $T_c$
cuprates, such as the emergence of superconductivity upon doping a
parent antiferromagnetic compound. \cite{chen,phasediagram,AFM}
Theoretical models \cite{theory1} suggest
unconventional $s^\pm$ pairing, consistent with Andreev
spectroscopy, \cite{pcs} penetration depth,
\cite{malone} and photoemission measurements
\cite{ding} which indicate nodless s-wave pairing symmetry. Experiments at high magnetic fields,
\cite{ding, hunte, weyeneth} penetration depth \cite{malone,hashimoto,uncon_lambda_122} and heat capacity measurements, \cite{keimer} are consistent with
multiband  $s^\pm$ scenarion\cite{theory1}
Other models suggest an important role for electronic correlations, \cite{theory2}
and even a possibility of unconventional pairing mechanisms. \cite{patrick}

The comparatively high critical temperatures $T_c$  and extremely high upper
critical fields $H_{c2}$ of the oxypnictides \cite{hunte,jo}
indicate their promising prospects for applications if,
unlike the layered cuprates, a sizeable vortex liquid region
does not dominate their temperature-magnetic field $(T-H)$ phase diagram. It is therefore
important to reveal the true behavior of the anisotropic
magnetization in the vortex state of the oxypnictides,
particularly the extent to which  vortex properties are affected
by strong magnetic correlations and multiband effects. For instance,
multiband effects in MgB$_2$ can manifest themselves in strong
temperature and field dependencies for the mass anisotropy
parameter $\gamma(T,H)=(m_c/m_{ab})^{1/2}$ and the London penetration depth
$\lambda(T,H)$ even at $H\ll H_{c2}$. \cite{MgB2vsT, MgB2vsH} Similar effects
in pnictides would be consistent with the multiband pairing scenarios. \cite{theory1}
Yet, there are significant differences between two-band
superconductivity in MgB$_2$ and in oxypnictides: in MgB$_2$ the
interband coupling is weak, while in the oxypnictides it is the
strong interband coupling which is expected to result in the high
$T_c$. \cite{theory1} Thus, probing multiband superconductivity in
oxypnictides by magnetization measurements requires high magnetic
fields, which can suppress the superfluid densities in both bands by circulating vortex currents.
In this work we address these issues, presenting
high-field torque measurements of the anisotropic reversible
magnetization in LaFeAsO$_{0.9}$F$_{0.1}$
single crystals. Our measurements of $\gamma(T,H)$ up to 18T and
extended temperature range, $4<T<15$ K reveals a different
behavior in $\gamma(T,H)$ as compared to recent low-field torque
measurements. \cite{weyeneth} Ref. \onlinecite{weyeneth}
shows a $\gamma$ that increases continuously as $T$ is lowered reaching a maximum value
of $\sim 20$ for both NdFeAsO$_{0.8}$F$_{0.2}$ and SmFeAsO$_{0.8}$F$_{0.2}$ single crystals. In contrast, our results indicate that $\gamma$ in LaFeAsO$_{0.9}$F$_{0.1}$ reaches a maximum of $\sim 10$ decreasing asymptotically towards 1 as $T$ is lowered.

Measurements of the equilibrium magnetization $m(T,H)$ of the vortex
lattice in LaO$_{0.9}$F$_{0.1}$FeAs are complicated by the smallness
of $m(T,H)$ caused by the large Ginzburg-Landau parameter,
$\kappa=\lambda/\xi > 100$ and by the background paramagnetism of
the normal state, \cite{para} which can mask the true behavior of
$m(T,H)$.  In this case torque magnetometry is the most sensitive
technique to measure the fundamental anisotropy of the parameters of
$\overrightarrow{m}(T,\overrightarrow{H})$ in small single crystals. The torque $ \overrightarrow{\tau}= \mu_0
\overrightarrow{m} \times \overrightarrow{H}$ acting upon a uniaxial superconductor is given by
    \begin{equation}
    \tau(\theta) = \frac{HV \phi_0 (\gamma^2-1)\sin2\theta}{16\pi\mu_0\lambda_{ab}^2\gamma^{1/3}\varepsilon(\theta)}
    \ln \left[ \frac{\eta H_{c2}^{ab}}{\varepsilon(\theta) H}\right]+\tau_m\sin 2\theta,
    \label{KoganEq}
    \end{equation}
where $V$ is the sample volume, $\phi_0$ is the flux quantum,
$H_{c2}^{ab}$ is the upper critical field along the ab planes,
$\eta\sim 1$ accounts for the structure of the vortex core, $\theta$
is the angle between $\overrightarrow{H}$ and the c-axis, $\varepsilon(\theta) =
(\sin^2 \theta+\gamma^2\cos^2\theta)^{1/2}$ and $ \gamma =
\lambda_c/ \lambda_{ab}$ is the ratio of the London penetration
depths along the c-axis and the ab-plane. The first term in Eq.
(\ref{KoganEq}) was derived by Kogan in the London approximation
valid at $H_{c1}\ll H\ll H_{c2}$, \cite{kogan} while the last term
describes the torque due to the background paramagnetism, for which
$\tau_m= \mu_0 (\chi_c-\chi_a)VH^2/2$ and $\chi_c$ and $\chi_a$ are the
normal state magnetic susceptibilities along the c-axis and ab-
plane, respectively. In general, $\gamma = \left(m_{c}/m_{ab}\right)^{1/2} \neq \lambda_c/ \lambda_{ab}$, (where $m_c$ and $m_{ab}$ are
the Ginzburg-Landau superconducting effective masses) but both ratios are assumed
to be equal in the model leading to Eq. \ref{KoganEq}.
As will be shown below, the paramagnetic term
in Eq. (\ref{Kogan}) in LaO$_{0.9}$F$_{0.1}$FeAs can be larger than
the superconducting torque, which makes extraction of the
equilibrium vortex magnetization nontrivial. In this work we propose
a method, which enables us to resolve this problem and measure the
true angular dependence of the superconducting torque as a function
of both $\overrightarrow{H}$ and $T$, probing the concomitant behavior of
$\gamma(T,H)$ and $\lambda_{ab}(T,H)$ and manifestations of
multiband effects.

\section{Experimental}
Underdoped single crystals of LaO$_{1-x}$F$_x$FeAs with typical
sizes of $80 \times 60\times 5 $ $\mu$m$^3$ were grown by the flux method described in Ref.
\onlinecite{zhigadlo}. The samples had a critical
temperature $T_c \simeq 15$ K as determined by the SQUID magnetometry, and as shown in Fig. 1.  The width of superconducting transition, measured by a commercial SQUID magnetometer under a field $H = 10$ Oe
after cooling the crystal under zero field, is $\Delta T_c \sim 3.5 $ K.
This relatively broad transition may not reflect the sample quality but mostly results from the
penetration of vortices in a plate-like crystal which has a large demagnetization factor and thus reduced lower critical field. Although in some crystals from the same batch the width of the resistive
transition, from the very onset of the resistive transition to the zero resistance state, is observed to be as large as $\Delta T \simeq 3$ to 4 K , consistent with the values reported in the literature for crystals with similar composition \cite{kim}. The fraction of F quoted here, corresponds to a nominal value since its precise content is very difficult to determine in such small single crystals. However, a superconducting transition temperature $T_c \simeq 15$ K, see Fig. 1, firmly places
these crystals within the underdoped state, following the overall phase diagram displaying $T_c(x)$ as a function
of the F content $x$. \cite{luetkens} As argued in Ref. \onlinecite{beek}, F is expected to be inhomogeneously distributed
throughout the samples.
\begin{figure}[htb]
\begin{center}
\epsfig{file= 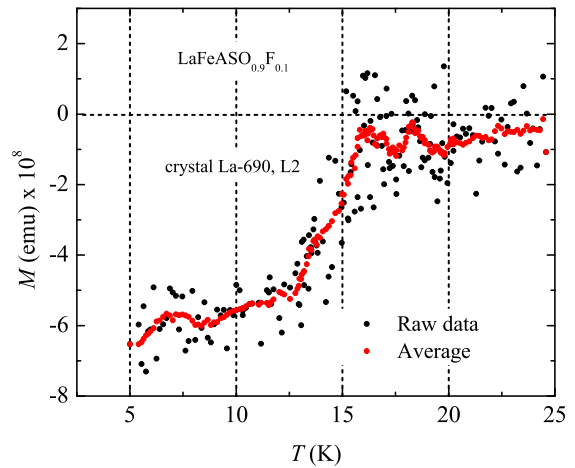, width = 7.4 cm} \caption {(color
online) (a) Magnetization $M$ as a function of temperature for a typical LaO$_{0.9}$F$_{0.1}$FeAs
single crystal measured under an external field of 10 Oe. The small mass of the single crystal, of just a few micrograms, leads
to the observed large scattering in the data points. Red markers correspond to $M(T)$ averaged over 20 raw data points.}
\end{center}
\end{figure}
Samples were attached to the tip of a
piezo-resistive micro-cantilever placed in a rotator inserted into a
$^3$He cryostat. The ensemble was placed into a 18 T superconducting
solenoid. Changes in the resistance of the micro-cantilever
associated with its deflection and thus a finite magnetic torque
$\tau$ was measured via a Wheatstone resistance bridge.

\section{Results and discussion}

\begin{figure*}[htb]
\begin{center}
\epsfig{file= 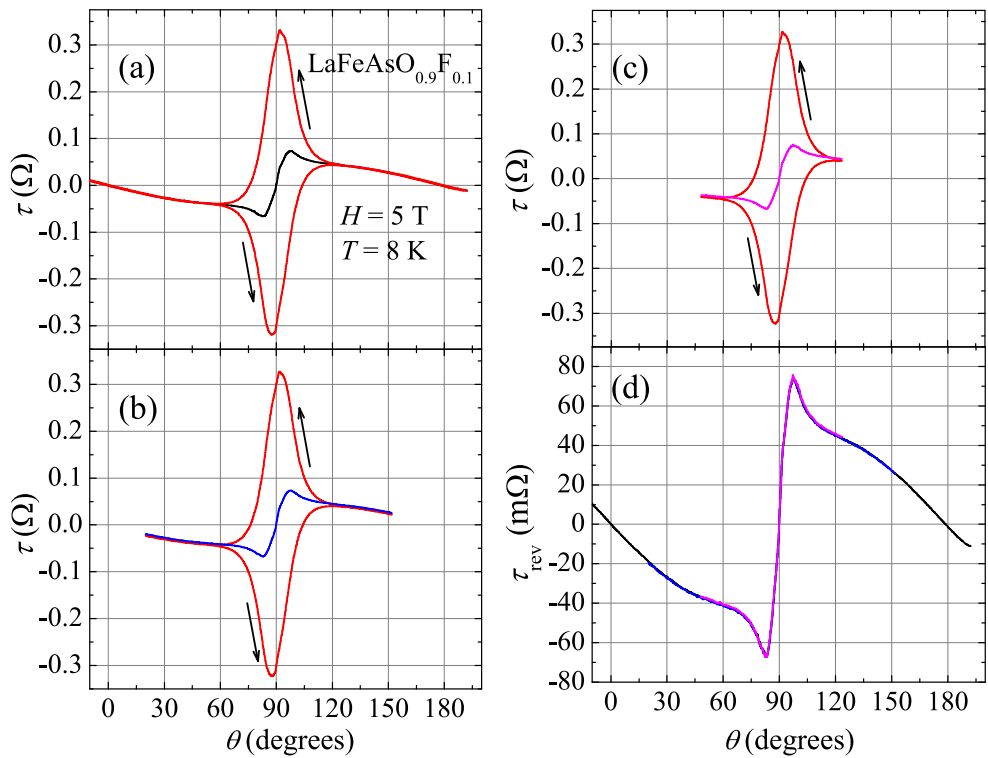, width = 13 cm} \caption {(color
online) (a) Magnetic torque $\tau$ for a LaO$_{0.9}$F$_{0.1}$FeAs
single crystal for increasing and decreasing angle ($\theta$) sweeps (red lines), at $H=5$ T and $T=8$ K.
Arrows indicate either increasing ($\theta_{up}$) or decreasing ($\theta_{down}$) angles.
Black line is the reversible torque component or $\tau_{\text{rev}}(\theta)$ defined here as the average between both traces.
(b) Same as in (a) but in a smaller angular range. (c) Same as in (a) and in (b) but in an even shorter angular range.
(d) The resulting $\tau_{\text{rev}}(\theta)$ from Figs. (a), (b) and (c).}
\end{center}
\end{figure*}

\begin{figure}[htb]
\begin{center}
\epsfig{file= 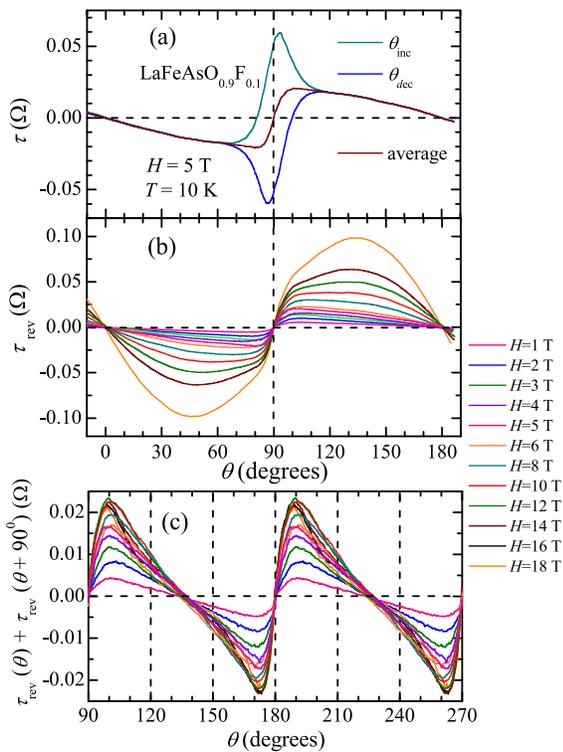, width = 7.4 cm} \caption {(color
online) (a) Magnetic torque $\tau$ for a LaO$_{0.9}$F$_{0.1}$FeAs
single crystal for increasing (clear blue line) and decreasing (blue
line) angle ($\theta$) sweeps, at $H=3$ T and $T=27$ K.
(b) $\tau_{\text{rev}}(\theta)$ for several field
values at $T = 10$ K. (c)
The symmetrized component $\tau_{\text{rev}}(\theta)+\tau_{\text{rev}}(\theta+90^{\circ})$ of
the torque due solely to the reversible vortex magnetization.}
\end{center}
\end{figure}
\begin{figure}[htb]
\begin{center}
\epsfig{file= 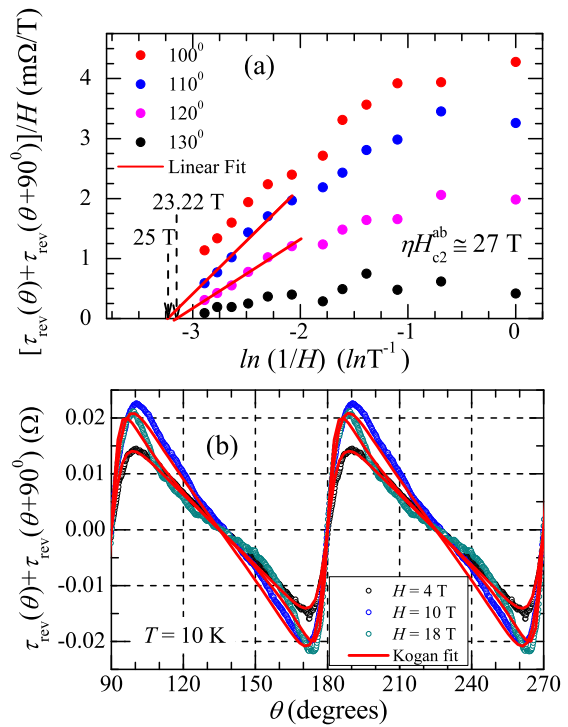, width = 7.4 cm} \caption {(color
online) (a) The amplitude of $[\tau_{\text{rev}}(\theta)+\tau_{\text{rev}}(\theta+90^{\circ})]/H$ as a function of $\ln (H^{-1})$ for several angles, taken from the data in Fig. 1 (c). Red lines are linear extrapolations of $[\tau_{\text{rev}}(\theta)+\tau_{\text{rev}}(\theta+90^{\circ})]/H \rightarrow 0$ for $\theta =110^{\circ}$ and $120^{\circ}$, which yields $\eta H_{c2}^{ab} \simeq 27 $ T for LaFeAsO$_{0.9}$F$_{0.1}$ at $T=10$ K. (b) With the value $\eta H_{c2}^{ab} \simeq 27 $ we obtain excellent fits of $\tau_{\text{rev}}(\theta)+\tau_{\text{rev}}(\theta+90^{\circ})$ to Eq. (2) for all field values.}
\end{center}
\end{figure}
\begin{figure}[htb]
\begin{center}
\epsfig{file= 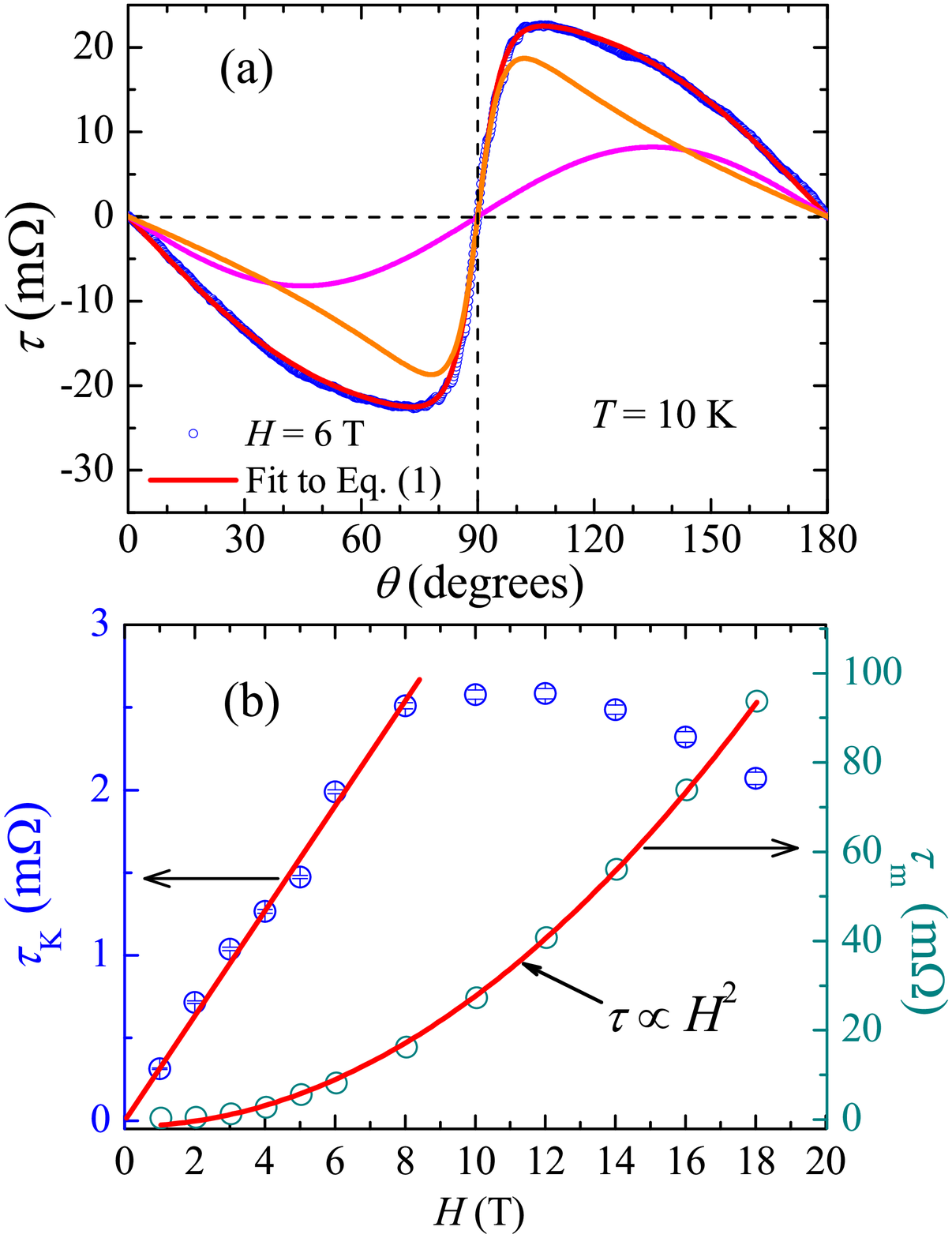, width = 7.4 cm} \caption {(color
online) (a) Reversible torque $\tau_{\text{rev}}(\theta)$ for a
LaFeAsO$_{0.9}$F$_{0.1}$ single crystal (blue markers) at $H = 6$ T and $T= 10$ K. Orange line shows the reversible vortex torque $\tau_K$ and the magenta line shows the paramagnetic torque extracted from the data shown in Fig. 3 using the deconvolution procedure described in the text. Red line shows the sum of $\tau_K$ with the paramagnetic term. (b) The amplitude of $\tau_K$ normalized by $(\gamma^2-1)/ \gamma^{1/3}$ (blue markers) and that of the paramagnetic term (clear blue markers) as functions of $H$. At lower fields, $\tau_K(H,\theta)$ increases linearly with $H$, while the paramagnetic term exhibits a $H^2$ dependence.}
\end{center}
\end{figure}

Fig. 2 (a) shows the angular dependence of the torque $\tau(\theta)$
at $T=8$ K and under $H=5$ T for a LaFeAsO$_{0.9}$F$_{0.1}$ single crystal.
Hysteresis, resulting from the irreversible
magnetization due to vortex pinning is observed between increasing
and decreasing angular sweeps. Black line depicts the average value of
both traces, $\tau_{\text{rev}}(\theta) = (\tau(\theta)_{\text{up}}+
\tau(\theta)_{\text{down}})/2$ defined as an equilibrium
magnetization, where $\theta_{\text{up}}$ and $\theta_{\text{down}}$
indicate either increasing or decreasing angle sweeps, respectively.
This reversible torque contains both superconducting and
paramagnetic contributions. Given that vortex pinning and concomitant
hysteresis are bound to disturb the equilibrium magnetization, we have checked
that our definition of $\tau_{\text{rev}}(\theta)$ leads to reproducible traces, by
re-measuring $\tau(\theta)$ for both $\theta_{\text{up}}$ and $\theta_{\text{down}}$ within
two distinct angular ranges, as shown in Figs. 2 (b) and 2 (c). The respective $\tau_{\text{rev}}(\theta)$
traces, i.e. blue and magenta traces, are plotted in Fig. 2 (d) together with the original $\tau_{\text{rev}}(\theta)$
from Fig. 2 (a). As seen, all traces overlap almost perfectly. Both sharp spikes seen at either side of $\theta = 90^{\circ}$
are likely due to pinning effects.

Fig. 3 (a) shows the angular dependence of the torque $\tau(\theta)$
at $T=10$ K and $H=5$ T for a \emph{second} LaFeAsO$_{0.9}$F$_{0.1}$ single crystal.
In this crystal and at this temperature one does not observe any structure that might be
attributable to pinning and which would compromise a fit attempt of the resulting $\tau_{\text{rev}}(\theta)$
(brown line) to either Eq. (1) or Eq. (2). As shown in Fig. 3(b), the paramagnetic term $\tau_m \sin 2\theta$ increases
rapidly as the field increases, preventing a direct fit of
$\tau_{\text{rev}}(\theta)$ to Eq. (1) since $\tau_m \sin 2\theta$ interferes
with the $\sin 2\theta$ harmonics of the first
term. Yet the  superconducting parameters $\gamma(T,H)$ and
$\lambda(T,H)$  can be unambiguously extracted from the data adding
a $90^{\circ}$ translation of $\tau_{\text{rev}}(\theta)$ to
itself, i.e. $\tau_{\text{rev}}(\theta)+\tau_{\text{rev}}(\theta +
90^{\circ})$, where the paramagnetic term in $\tau(\theta)$ cancels out:
    \begin{eqnarray}\label{Kogan}
    \tau_+=\tau(\theta)+ \tau(\theta + 90^{\circ}) = \frac{V \phi_0(\gamma^2-1) H\sin2 \theta}{16\pi \mu_0 \lambda_{ab}^2\gamma^{1/3}}
    \nonumber \\
    \times \left[ \frac{1}{\varepsilon(\theta)} \ln \left(
    \frac{\eta H_{c2}^{ab}}{\varepsilon(\theta) H}
    \right) - \frac{1}{\varepsilon^{\star}(\theta)} \ln \left( \frac{\eta H_{c2}^{ab}}{\varepsilon^{\star}(\theta) H}
    \right) \right],
    \end{eqnarray}
where $\varepsilon^{\star}(\theta) = ( \cos^2\theta + \gamma^2
\sin^2\theta )^{1/2}$. This procedure is illustrated by Fig. 3 (c) which shows $\tau_+(\theta)$ as a function of $\theta$
for all curves in Fig. 3 (b). Notice that the amplitude of $\tau_+(\theta)$ is considerable smaller than that of $\tau_{\text{rev}}(\theta)$.

Since $\tau_+(\theta)$ in Eq. (2) depends on three fit
parameters, $\gamma$, $\lambda$ and $\eta H_{c2}$, different sets of
parameters may give equally good descriptions of the
experimental data. One can circumvent this difficulty by extracting
$\eta H_{c2}$ from the amplitude of $\tau_+(\theta)/H$ plotted as a function of $\ln(1/H)$ for several values of
$\theta$, as shown in Fig. 4 (a). If at a given temperature, these
measurements are performed up to high enough fields, the
extrapolation of $\tau_+(\theta)/H$ to zero evaluates the value
$H_{\star}$ at which $\left(\eta H_{c2}^{ab} / \varepsilon(\theta)
H_{\star} \right) = 1$. From the two extrapolated values of
$H_{\star}$ for $\theta_1 = 110^{\circ} $ and
$\theta_2=120^{\circ}$, we exclude $\gamma$ and obtain $(\eta
H_{c2})^2 =
H_{\star}^2(\theta_1)H_{\star}^2(\theta_2)\sin(\theta_2-\theta_1)\sin(\theta_2+\theta_1)
/[H_{\star}^2(\theta_1)\cos^2\theta_1-H_{\star}^2(\theta_2)\cos^2\theta_1]$,
which yields  $\eta H_{c2}^{ab} \simeq 27$ T for $T = 10 $ K. This
simple method provides a \emph{thermodynamic} estimate for
$H_{c2}^{ab}$ without the need of extremely high magnetic fields,
$H\simeq H_{c2}(\theta)$. Furthermore, with $\eta H_{c2}^{ab} \simeq
27$ T we obtain excellent and stable two-parameter fits of
$\tau_+(\theta)$ to Eq. (2) for virtually all field values, as shown
in Fig. 4 (b). Here the parameters $\gamma$ and $\lambda$ no longer
interfere in the fit as $\lambda$ affects only the magnitude but not the shape of
$\tau_+(\theta)$. We confirmed that the extracted parameters lead to
a small difference in the logarithmic terms of Eq. (2) at $H= H_{\star}$.

Using the method outlined above, we extracted the field and the temperature
dependencies of $\gamma(T,H)$, $\lambda_{ab} (T, H)$. An example, is given by the orange line
in Fig. 5 (a) for $H = 6$ T and $T = 10$ K. Having fixed $\gamma$, $\lambda$, and $\eta H_{c2}^{ab}$ we can now fit the original
$\tau_{\text{rev}}(\theta)$ to Eq. (1) (red line), leaving  the
amplitude of the paramagnetic component $\tau_m$ as the only
adjustable parameter (magenta line). Figure 5 (b) depicts the
resulting amplitudes for both $\tau_m$ and the superconducting
contribution $\tau_K \propto H / \lambda_{ab}^2$ in Eq. (1) as a
function of field for $T = 10$ K. The amplitude of $\tau_K$ follows
the expected linear in field dependence up to $H = 8$ T, from which
point it starts to decline continuously. On the other hand, $\tau_m$
follows the expected $H^2$ dependence characteristic of the torque
of an anisotropic paramagnetic background. To our knowledge, this is
the first magnetometry method which allows an independent extraction
of the parameters of a magnetic superconductor.

The curves $\tau(\theta)$ measured for $T> 4$ K, indicate that $\eta
H_{c2}^{ab}(T)$ is described by $\eta H_{c2}^{ab}(T) = \eta
H_{c2}^{ab}(0) [1-(T/T_c)^2]$. As seen in Fig. 5, the irreversible component in $\tau(\theta)$ grows quickly as the temperature is lowered.
At $T = 4$ K despite the observed large irreversibility, the resulting reversible component in $\tau(\theta)$ is still nearly perfectly described by Eq. (1),
see Fig. 6 (a). However, as the temperature is lowered to $T = 1.5$ K additional structures emerge in $\tau(\theta)$ for $\theta$ close to $90^{\circ}$, see Fig. 6 (b).
We ascribe these features to the intrinsic pinning of vortices by the planar structure of the material, which prevents a reliable extraction of the reversible
component in $\tau(\theta)$ below $T \sim 4$ K. A detailed analysis and discussion will be provided elsewhere. \cite{gang}
\begin{figure}[htb]
\begin{center}
\epsfig{file= 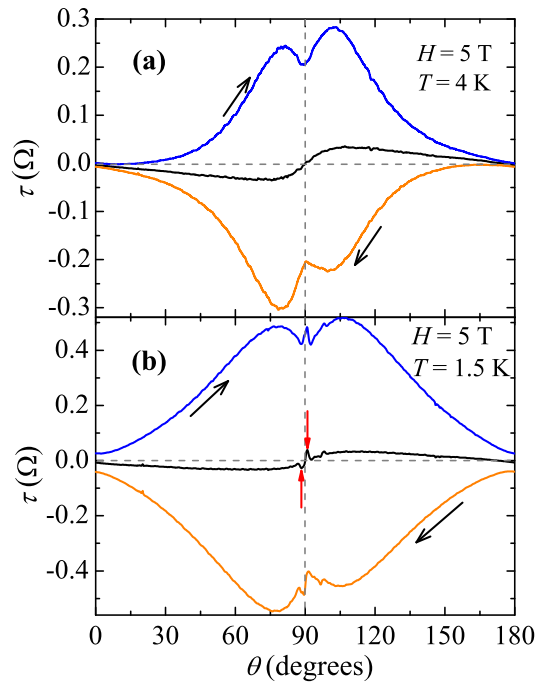, width = 7 cm} \caption {(color
online) (a) $\tau$ as a function of the angle $\theta$ and respectively for increasing (blue line) and decreasing (orange line) angle scans at $T = 4$ K and under a field $H = 5$ T. Black line corresponds to the average between both traces. (b) Same as in (a) but for $T = 1.5$ K. Notice the emergence of sharp peaks in $\tau(\theta)$ (red arrows) for $\theta$ close to $90^{\circ}$ resulting from the intrinsic pinning by the planar structure.}
\end{center}
\end{figure}
\begin{figure}[htb]
\begin{center}
\epsfig{file= 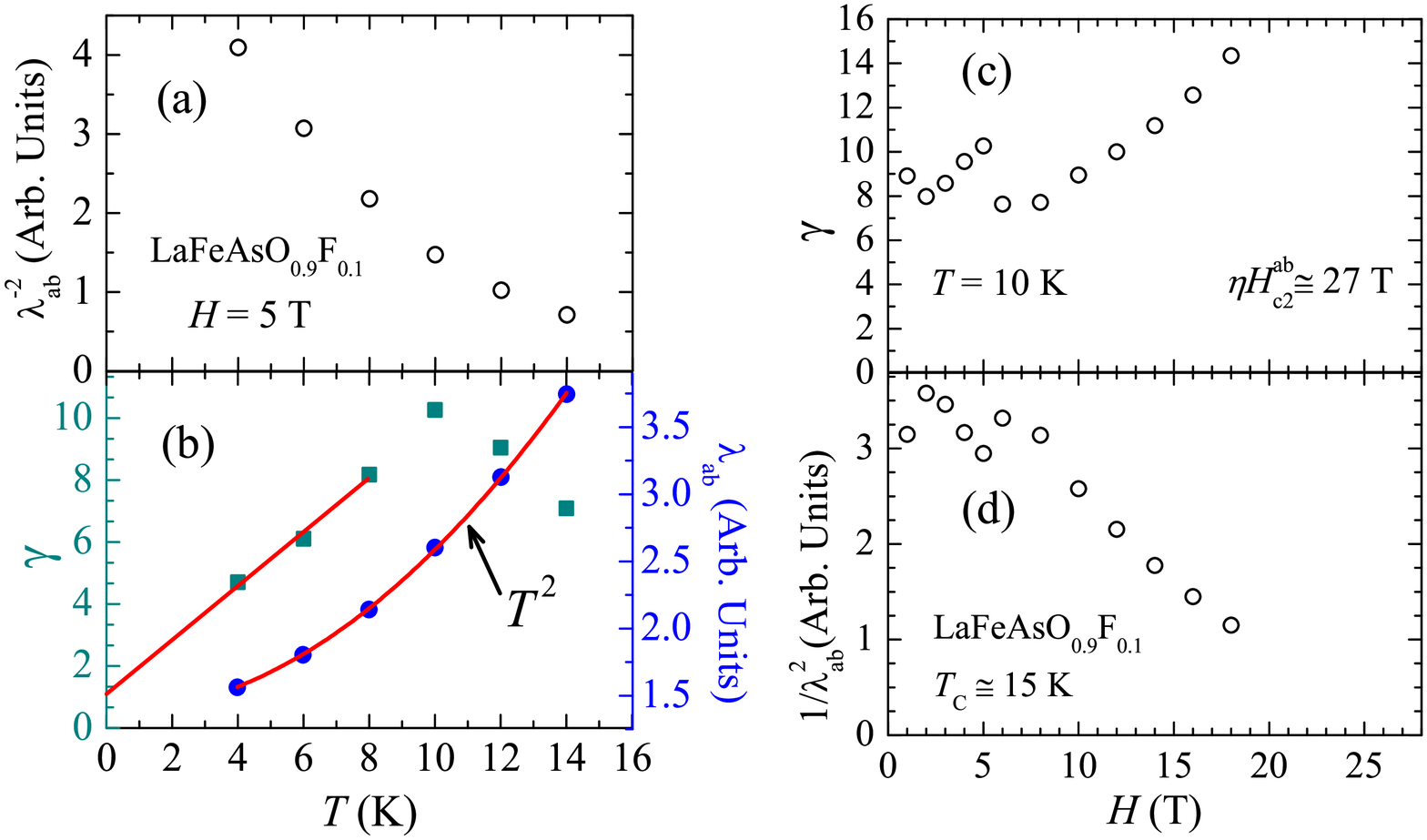, width = 8.7 cm} \caption {(color
online) (a) Temperature dependence of the superfluid density $n_s \propto \lambda^{-2}_{ab}$ extracted from the torque data at $H=5$ T. (b) Temperature dependence of both $\gamma$ (clear blue markers) and $\lambda_{ab}$ (blue markers). Here  $\lambda_{ab}(T)$ exhibits a quadratic dependence on $T$. (c) $\gamma$ as a function of magnetic field as extracted from the fittings in Fig. 2 (b). (d) The superfluid density $n_s \propto \lambda^{-2}_{ab}$ as a function of $H$ at $T = 10$ K. }
\end{center}
\end{figure}

Figure 7 (a) shows the resulting
temperature dependence of the total superfluid density $n_s \propto
\lambda_{ab}^{-2}$ measured under a field $H = 5$ T. A field of 5 T is high enough to suppress the irreversible component in the torque as compared to the reversible one, and to neglect demagnetization factors and geometrical barriers for the penetration of magnetic flux through the sample.
Our torque magnetometer was not calibrated, so we could only measure the temperature dependency of $\lambda_{ab}(T)$ but not
the absolute values of $\lambda_{ab}$ or $n_s$. Despite the
unavailability of points at the lowest $T$s, the temperature dependence of $n_s(T)$ exhibits
a positive curvature as seen in penetration depth measurements in the compounds \emph{R}FeAsO$_{0.9}$F$_{0.1}$ (\emph{R}=La, Nd, with $T_c$s of 14 and 45 K, respectively). \cite{uncon_lambda} But it contrasts with penetration depth results for PrFeAsO$_{1-y}$ ($T_c \simeq 35$ K) \cite{con_lambda} and for SmFeAsO$_{0.8}$F$_{0.2}$ ($T_c \simeq 44$) K \cite{malone}, which
finds evidence for a superconducting state characterized by more than one (non-nodal) gap.
In Ref. \onlinecite{uncon_lambda} such an upward curvature was explained in terms
of the $s^\pm$ scenario, with the superconducting gap ratio $\Delta_1 / \Delta_2
\simeq 1/3$.  As follows from Fig. 7 (b), the associated $\lambda_{ab}(T)$ in our limited temperature range
follows a  $T^2$ dependence at low and intermediate temperatures, in contrast to the linear
dependence expected for a nodal gap as observed in LaFePO
\cite{Tlinear}, or the nearly exponential behavior expected for a
clean s-wave superconductor, as reported for
SmFeAsO$_{0.8}$F$_{0.2}$ \cite{malone} and PrFeAsO$_{1-y}$.
\cite{con_lambda} One important debate concerning the Fe arsenides
is precisely the origin of these differences, since a common
pairing mechanism \cite{theory1,theory2} has been proposed for all compounds based on
the presumed similarity of their electronic structure.
At the same time, it was recently argued that the extended
$s$-wave scenario can lead to either nodeless or nodal gaps,
depending on the interplay between intraband and interband interactions. \cite{chubukov}
Penetration depth measurements over an extended
region in reduced temperatures $(T/T_c)$ also revealed a
$T^2$-dependence for $\lambda_{ab}$ in $R$FeAsO$_{0.9}$F$_{0.1}$
($R=$La, Nd) \cite{uncon_lambda} as well as in Ba$_{1-x}$K$_x$Fe$_2$As$_2$.
\cite{uncon_lambda_122} This power-law temperature dependence was
deemed to be consistent with the $s \pm$ muti-gap scenario, if
either strong interband impurity scattering \cite{vorontsov} or
pair-breaking effects \cite{pair_braking} are important.
Our observation of a $n_s \propto T^2$  is limited
by the restricted temperature range imposed by the pinning effects at lower $T$s, yet
we believe that it is important to expose the agreement for the anomalous temperature dependence of the penetration depth between magnetic torque and surface impedance measurements in the case of LaFeAsO$_{0.9}$F$_{0.1}$. Given the width of the superconducting transition $\simeq 3-4$ K, it is possible that local variations of $T_c$ due to an inhomogeneous distribution of F, could affect the dependence of $n_s$ on $T$ at higher temperatures. In addition, given that our measurements of $n_s(T)$ were performed under a field of 5 T, we cannot completely rule out that the temperature dependence of $H_{c2}(T)$ might influence the power law dependence for $n_s(T)$ reported here, particularly at higher temperatures.


Figure 7 (b) also displays the
temperature dependence of the mass anisotropy parameter $\gamma$,
which starts at $\gamma\simeq 7$ for $T \leq T_c$, increases toward
a maximum of $ \sim 10$ at $T \simeq 0.75$ $T_c$,  and then decreases
to  $\approx 1$ as $T$ is lowered (as it happens in nearly all Fe based
superconductors, see Refs. \onlinecite{hunte,jo}). Our unpublished
transport measurements in LaFe$_{1-x}$Co$_x$AsO indicate a very
similar $T$-dependence for $\gamma_H = H_{c2}^{ab}/H_{c2}^c$.

Figures 7 (c) and 7 (d) show the field dependencies of $\gamma$ and
$n_s$ for $T = 10$ K. Here $\gamma$  increases by nearly a factor of
2, from $ \sim 8$ at small fields to $\sim 15$ at $H=18$ T. On the
other hand, $n_s$ remains nearly  constant in fields up to $H_v = 8$
T, and then decreases at higher fields, extrapolating to $n_s=0$ at
$ H(\text{10 K}) \approx 25 $ T.  The complete suppression of $n_s$ at this value is consistent
with our previous estimate of $\eta H_{c2}\approx 27$ T obtained above.
The suppression of $n_s(H)$ upon application of magnetic field is consistent with the onset of orbital pair-breaking by
circulating vortex currents as the spacing between the vortex cores
becomes smaller than $\approx \sqrt{3}$ of the vortex spacing at
$H_{c2}$ where superconductivity is fully suppressed. For instance, the single-band
Ginzburg-Landau theory predicts $\lambda^{-2}\propto n_s(H) \simeq n_s(0) (1-H/H_{c2}) $ for $T$ close to $T_c$. \cite{brandt}
Interestingly, the ratio  $H_v/H_{c2}\simeq 1/3$ turned out to be approximately
equal to the ratio of the superconducting gaps on the electron and
hole pockets of the Fermi surface as observed in Ref.
\onlinecite{uncon_lambda}. Around $ H = $ 6 - 7 T, there is a spike in
$\lambda$ and/or a dip in $n_s$, the origin of which remains unclear.

We described our data using the simplest Eq. (1) for the torque in a
uniaxial superconductor, assuming that the angular dependencies of
$\lambda$ and $\xi$ are controlled by the single anisotropy
parameter $\gamma$. A more complicated expression for $\tau$ with
two different anisotropy parameters
$\gamma_H=H_{c2}^{ab}/H_{c2}^{c}$ and $\gamma_{\lambda} =
\lambda_{ab} / \lambda_c$ was suggested for multiband
superconductors, \cite{koganprl} However given that the formula for
$\tau$ of Ref. \onlinecite{koganprl} was obtained on phenomenological
grounds, and the angular dependencies of $\lambda(\theta)$ and
$H_{c2}(\theta)$ derived from the multiband BCS theory are much more
complicated, \cite{ag} we believe that the use of that expression for
$\tau$ with different $\gamma_H$ and $\gamma_\lambda$ may not
unambiguously reveal any new physics as  compared to the simpler Eq.
(\ref{KoganEq}). Indeed, the fit of our data with $\tau$ from Ref.
\onlinecite{koganprl} gave behaviors qualitatively similar to those shown
above: $\gamma_H \simeq 10 \mid_{T \approx T_c}$ increases to a
maximum value of 18 before decreasing to much lower values as $T
\rightarrow 0$. Meanwhile  $\gamma_{\lambda} = 1.4 \mid_{T \approx
T_c}$ exhibits a mild decrease towards 1 as $T \rightarrow 0$. This
common tendency of $\gamma$ to approach a value close to 1 upon
decreasing $T$, observed on many Fe-based superconductors, indicates
the increasing contribution of the Pauli pair-breaking effects at lower $T$s
which is neither taken into account in Eq. (1) nor in the expression
for $\tau$ of Ref. \onlinecite{koganprl} .

In conclusion, we report the torque measurements of reversible
magnetization of LaO$_{0.9}$F$_{0.1}$FeAs single crystals from which
the field dependent mass anisotropy parameter $\gamma$ and the
reduced London penetration depth $\lambda(H,T)/\lambda(0,T)$ were
measured. The significant field dependencies of $\gamma $ and
$\lambda(H,T)/\lambda(0,T)$ are consistent with multiband pairing
while the observed quadratic temperature  dependence of
$\lambda(H,T)/\lambda(0,T)$ at low and intermediate $T$ is consistent with
the $s^\pm$ pairing with interband impurity scattering.

\section{Acknowledgements}

We acknowledge S. Weyeneth for the SQUID measurements. The NHMFL is supported by NSF through NSF-DMR-0084173 and the State of Florida.  L.~B. is supported by DOE-BES through award DE-SC0002613.
Work at ETH was supported by the Swiss National Science Foundation and the NCCR program MaNEP.

\end{document}